\documentclass[12pt]{article}

\begin{document}

\title{General Theorems on Decoherence in the Thermodynamic Limit\footnote{e-mail:marcofrasca@mclink.it}}

\author{Marco Frasca \\
Via Erasmo Gattamelata, 3,\\
         00176 Roma (Italy)}

\date{\today}

\maketitle

\abstract{
We extend the results on decoherence in the thermodynamic limit 
[M. Frasca, Phys. Lett. A {\bf 283}, 271 (2001)]
to general Hamiltonians. It is shown that N independent particles, initially properly
prepared, have a set of observables behaving classically in the thermodynamic limit.
This particular set of observables is then coupled to a quantum system that in this
way decoheres so to have the density matrix in a mixed form. This gives
a proof of the generality of this effect.
}

PACS: 03.65.Yz, 5.70.Ln

\newpage

Decoherence \cite{zur} appears as a rather ubiquitous effect in quantum systems. Due to
the interaction with a large environment, a quantum system tends to
modify its quantum evolution from unitary, that means coherent, to a decaying
form. In this terms, we can say that decoherence, as commonly understood, is
a dissipative effect. Then, the density
matrix containing interference terms, after a trace procedure on the environment
degrees of freedom gets a mixed form so to have
probabilities attributed to the outcome of possible measurements on the system. 
As a matter of fact, this does not solve the
measurement problem in quantum mechanics \cite{cri}. The main question is that,
after a measurement, one gets a pure state for the quantum system  and not just
the mixed form of the density matrix that is obtained by decoherence.

Anyhow, the appearance of decaying effects on unitary evolution of a quantum system
is generally seen in experiments and one can safely affirm that decoherence is
a rather well verified phenomenon \cite{exp}. In this paper we want to go one
step beyond on the basis of a recent proposal for non dissipative decoherence
in the the thermodynamic limit \cite{fra1}. We will prove that a pure state is
indeed obtained when a large number of quantum systems interacts with another one,
washing out superposition states and approaching in this way a possible solution
to the measurement problem, for a single event, self consistently inside quantum mechanics. 
The essential
characteristic of our approach is that unitary evolution is preserved and decoherence is
dynamically produced. 

First of all, we will prove a general theorem on the appearance of classical
states in the dynamical evolution of an ensemble of non interacting quantum
systems. We show that:

\newtheorem{classical}{Theorem}
\begin{classical}[Classicality]
An ensemble of N non interacting quantum systems, for properly chosen initial
states, has a set of operators $\{A_i,B_i,\ldots:i=1\ldots N\}$ from which one can derive a set
of observables behaving classically.
\end{classical} 

To prove this theorem we consider a Hamiltonian of the form
\begin{equation}
    H = \sum_{i=1}^N H_i \label{eq:H}
\end{equation}
and a set of observables $\{A_i,B_i,\ldots:i=1\ldots N\}$. For each system we take a set of
states of single particle  $|\psi_i\rangle$ in a such a way to have the initial product state
\begin{equation}
    |\psi(0)\rangle = \prod_{i=1}^N|\psi_i\rangle \label{eq:psi}
\end{equation}
not being an eigenstate of the Hamiltonian (\ref{eq:H}). Firstly, we prove that
the Hamiltonians $H_i$ belong to the set of operators $\{A_i,B_i,\ldots:i=1\ldots N\}$, 
being the corresponding observable the Hamiltonian $H$. This means that the
mean value of $H$ on the state (\ref{eq:psi}) is overwhelming large with respect to
its quantum fluctuations in the thermodynamic limit. This proves that the set
$\{A_i,B_i,\ldots:i=1\ldots N\}$ is not empty. Indeed, by straightforward algebra one has
\begin{equation}
    \langle H\rangle = N\bar H
\end{equation}
having set
\begin{equation}
    \bar H = \frac{1}{N}\sum_{i=1}^N \langle\psi_i|H_i|\psi_i\rangle
\end{equation}
that is an average of the mean values of each $H_i$. This proves that the
mean value of $H$ is proportional to $N$. Without much more difficulty we get 
\begin{equation}
    \langle H^2\rangle = N\overline{H^2} + \sum_{i\neq j}\langle\psi(0)|H_iH_j|\psi(0)\rangle
	 \label{eq:hz}
\end{equation}
being
\begin{equation}
    \overline{H^2} = \frac{1}{N}\sum_{i=1}^N \langle\psi_i|H_i^2|\psi_i\rangle
\end{equation}
again an average of the mean values of the square of each $H_i$. We note at this stage
that the last term in eq.(\ref{eq:hz}) gives a contribution also proportional to $N^2$.
This gives at last the fluctuation
\begin{equation}
    (\Delta H)^2 = \langle H^2\rangle - \langle H\rangle^2 = N \overline{(\Delta H)^2},
\end{equation} 
the mean value of $H$ now removes the $N^2$ term, with 
\begin{equation}
    \overline{(\Delta H)^2} = \frac{1}{N}\sum_{i=1}^N 
	[\langle\psi_i|H_i^2|\psi_i\rangle-\langle\psi_i|H_i|\psi_i\rangle^2]
\end{equation}
showing that the square of the fluctuation of $H$ is proportional to the average
of the fluctuations of each $H_i$. So, one has that the mean value of $H$ is
proportional to $N$ while the fluctuation is proportional to $\sqrt{N}$ and then
the former is overwhelming large with respect to the latter, if the initial
Hamiltonian (\ref{eq:H}) has a large ensemble of systems composing it (thermodynamic
limit). Being the quantum fluctuations negligible, one can say that the quantum system
we are considering, in the thermodynamic limit, behaves classically with respect to the
observable $H$, if properly prepared in the state $|\psi(0)\rangle$. 
So, the ensemble $\{A_i,B_i,\ldots:i=1\ldots N\}$ is not empty.

Now, we extend the proof to any other operator that can belong to the set
$\{A_i,B_i,\ldots:i=1\ldots N\}$. If, for a given $i$, $A_i$ commutes with $H$, it does not evolve in time, being
a conserved observable, and the above argument for $H$ applies straightforwardly.
Instead, if, generally,  $[A_i,H]\neq 0$ we have to study the time evolution of these
observables by the Heisenberg equations of motion. By analogy with $H$ we introduce
the observable $A=\sum_{i=1}^N A_i$ (the same can be done with any other set of
operators belonging to the given set), then (here and in the following $\hbar = 1$)
\begin{equation}
    A(t)=e^{iHt}Ae^{-iHt}.
\end{equation}
With this definition it is not difficult to obtain, with the same state initial state,
\begin{equation}
    \langle A(t)\rangle = N \overline{A(t)}
\end{equation}
being
\begin{equation}
    \overline{A(t)} = \frac{1}{N}\sum_{k=1}^N\langle\phi_k|e^{iH_kt}A_ke^{-iH_kt}|\phi_k\rangle,
\end{equation}
and
\begin{equation}
    [\Delta A(t)]^2 = \langle A(t)^2\rangle - \langle A(t)\rangle^2 = N \overline{\Delta A(t)^2}
\end{equation}
being
\begin{equation}
    \overline{\Delta A(t)^2} = \frac{1}{N}\sum_{k=1}^N[\langle\phi_k|e^{iH_kt}A_k^2e^{-iH_kt}|\phi_k\rangle
	-\langle\phi_k|e^{iH_kt}A_ke^{-iH_kt}|\phi_k\rangle^2],
\end{equation}
proving finally the theorem as we have the fluctuation proportional to $\sqrt{N}$ and
the mean value proportional to $N$. We can recognize from this result a strict
similarity with statistical mechanics as it should be expected from the start (see \cite{kad}).

Once we have such a set of observables, one may ask if these operators can indeed
produce decoherence. This is the content of the next theorem:

\begin{classical}[Decoherence]
An ensemble of N non interacting quantum systems,
having a set of operators $\{A_i,B_i,\ldots:i=1\ldots N\}$, from which one can derive a set
of observables behaving classically, and strongly interacting with a quantum system
through a Hamiltonian having forms like $V_0 \otimes\sum_{i=1}^N A_i$, can produce
decoherence if properly initially prepared.
\end{classical} 

By ``strongly interacting'' we mean that the Hamiltonian of the $N$ non interacting
quantum systems can be neglected, and perturbation theory can be applied. We
want to use a theorem for strong coupling proved in Ref.\cite{fra2}. In fact,
the Hamiltonian of this system can be written, choosing as a observable
$\sum_{i=1}^N A_i$ acting in the Hilbert space of the bath,
\begin{equation}
    H_{SB} = H_S + \sum_{i=1}^N H_i + V_0 \otimes\sum_{i=1}^N A_i
\end{equation}
being $H_S$ the Hamiltonian of the quantum system, and $V_0$ an operator,
acting in the Hilbert space of the system,
coupling the quantum system to the bath of $N$ non interacting systems. So,
if we assume the coupling between the system and the bath to be very large, we
can apply the theorem of Ref.\cite{fra2} to the Hamiltonian in the interaction
picture in the system's variables
\begin{equation}
    H_I = e^{iH_S t}V_0e^{-iH_S t} \sum_{i=1}^N A_i,
\end{equation}
stating that the strong coupling approximation is given by
\begin{equation}
    |\psi(t)\rangle\approx\sum_ne^{i\dot\gamma_n t} e^{-iN\overline{a}v_nt}
	|v_n\rangle\langle v_n|\psi_S(0)\rangle\prod_{i=1}^N |\chi_i\rangle \label{eq:state}
\end{equation}
being $|\psi_S(0)\rangle$ the initial state of the quantum system, 
\begin{equation}
    V_0|v_n\rangle = v_n|v_n\rangle,
\end{equation}
assuming a discrete spectrum and
\begin{equation}
    \dot\gamma_n = \langle v_n|H_S| v_n\rangle.
\end{equation}
The initial state of the bath is chosen in such a way to have 
$A_i|\chi_i\rangle = a_i |\chi_i\rangle$ and being $\overline{a}=\sum_{i=1}^Na_i/N$
a constant.

The state (\ref{eq:state}) has a quite interesting aspect as the phases of the
oscillating exponentials, already at very small energy, can have a time scale
of the order of the Planck time in the thermodynamic limit, making senseless
the possibility to observe such oscillations on the corresponding probabilities.
This means, mathematically, that such 
probability oscillations are averaged away \cite{fra3}.
Indeed, for the density matrix of the system one has
\begin{eqnarray}
    \rho_S(t) &=& \sum_n |\langle v_n|\psi_S(0)\rangle|^2 |v_n\rangle\langle v_n| \\ \nonumber
	&+&\sum_{m\neq n} e^{i[\dot\gamma_m - \dot\gamma_n] t} e^{-iN\overline{a}[v_m - v_n]t}
	\langle v_m|\psi_S(0)\rangle\langle\psi_S(0)|v_n\rangle |v_m\rangle\langle v_n|
\end{eqnarray}
with the interference terms being averaged away on a maximum time scale 
$\tau_M = 1/(N\overline{a}\min[v_m - v_n])$ being $\min[v_m - v_n]$
the minimal energy difference between the eigenavalues of $V_0$. This time is
really small already at energies of order of eV for $N$ becoming very large and
generally comparable with the Planck time. As a final comment about this theorem,
we point out that use has been made of a strong coupling between a bath and a
system. This kind of coupling is not generally common but can be realized
in ion traps and could turn out to be used in quantum computation.

Finally, we prove a general theorem in measure theory in quantum mechanics,
stating that
\begin{classical}[Measure]
If the operator $V_0$, strongly coupled to a quantum system with an observable of
the ensemble of $N$ non interacting systems, is linear in the generators of coherent states,
Schr\"odinger cat states are washed out in the leading order of the coupling
in the thermodynamic limit.
\end{classical}

We assume, initially, that the system has the following Hamiltonian, neglecting
the bath contribution at the leading order in the strong coupling expansion,
\begin{equation}
    H_F = \omega a^\dagger a + (\gamma a^\dagger + \gamma^* a )\otimes\sum_{i=1}^N A_i
\end{equation} 
so that, we have $V_0$ given by a linear combination of generators of coherent states \cite{cs}.
Besides, the initial state is taken to be a superposition state as
\begin{equation}
    |\psi(0)\rangle = {\cal N}(|\alpha e^{i\phi}\rangle+|\alpha e^{-i\phi}\rangle)
	\prod_{i=1}^N |\chi_i\rangle
\end{equation}
being ${\cal N}$ a normalization factor and $|\alpha e^{\pm i\phi}\rangle$
coherent states as to have a Schr\"odinger cat state \cite{Schl}. 
At the leading order, we can write the unitary evolution operator as \cite{tm},
\begin{equation}
    U_F(t) = e^{i\xi(t)}e^{-i\omega a^\dagger at}\exp[\hat\beta(t)a^\dagger - \hat\beta(t)^* a]
\end{equation}
being
\begin{equation}
\hat\xi(t)=\frac{\left(\sum_{i=1}^N A_i\right)^2|\gamma|^2}{\omega^2}(\omega t - \sin(\omega t))
\end{equation}
and
\begin{equation}
\hat\beta(t)=\frac{\left(\sum_{i=1}^N A_i\right)\gamma}{\omega}(1 - e^{i\omega t}).
\end{equation}
So, it is straightforward to obtain the wave function as
\begin{eqnarray}
     |\psi(t)\rangle&\approx& U_F(t)|\psi(0)\rangle = e^{i\xi(t)}
	 {\cal N}\left(e^{i\phi_1(t)}|\beta(t)e^{-i\omega t} 
	 + \alpha e^{i\phi-i\omega t}\rangle\right. \\ \nonumber
	 &+& \left.e^{i\phi_2(t)}|\beta(t)e^{-i\omega t} + \alpha e^{-i\phi-i\omega t}\rangle\right)
	 \prod_{i=1}^N |\chi_i\rangle \label{eq:st}
\end{eqnarray}
being now
\begin{equation}
\xi(t)=\frac{N^2|\gamma|^2}{\omega^2}(\omega t - \sin(\omega t)),
\end{equation}
and
\begin{equation}
\beta(t)=\frac{N\gamma}{\omega}(1 - e^{i\omega t}).
\end{equation}
and
\begin{equation}
    \phi_1(t) = -i\frac{\alpha}{2}[\beta(t)e^{-i\phi}-\beta^*(t)e^{i\phi}],
\end{equation}
\begin{equation}
    \phi_2(t) = -i\frac{\alpha}{2}[\beta(t)e^{i\phi}-\beta^*(t)e^{-i\phi}].
\end{equation}
The state (\ref{eq:st}), in the thermodynamic limit $N\rightarrow\infty$, reduces
to a pure coherent state for the system, $|\beta(t)e^{-i\omega t}\rangle$, proving
the main assertion of the theorem. The Schr\"odinger cat state is washed away in
the thermodynamic limit. 

To keep this letter shorter, we omit to prove that the system of the last theorem
undergoes true decoherence. This can be seen from the interference term of the
Wigner function that, in the thermodynamic limit, displays very rapid oscillations
on a time scale of the Planck time or smaller, being not physical. We can apply
the theory of divergent series \cite{hardy} to assume in the sense of Abel or Euler that 
$\lim_{N\rightarrow\infty}\cos(Nf(t)) = 0$
and $\lim_{N\rightarrow\infty}\sin(Nf(t)) = 0$ and all boils down to an average
in time. So, no interference term can be actually observed and true classical
behavior emerges. This argument is similar to the one applied in the
proof of the second theorem.

In conclusion, we have generalized our approach to the study of quantum
mechanics in the thermodynamic limit, given in \cite{fra1}, to a large
class of quantum systems, proving its wide applicability.

\end{document}